# Identification of electronic dimensionality reduction in semiconductor quantum well structures

Takahito Takeda[1,†], Kengo Takase[1], Vladimir N. Strocov[2], Masaaki Tanaka[1,3], and Masaki Kobayashi[1,3,*]

[1]*Department of Electrical Engineering and Information Systems, The University of Tokyo, 7-3-1 Hongo, Bunkyo-ku, Tokyo 113-8656, Japan*
[2]*Swiss Light Source, Paul Scherrer Institut, CH-5232 Villigen PSI, Switzerland*
[3]*Center for Spintronics Research Network, The University of Tokyo, 7-3-1 Hongo, Bunkyo-ku, Tokyo 113-8656, Japan*

Authors to whom all correspondence should be addressed: [†]ttakeda@g.ecc.u-tokyo.ac.jp, [*]masaki.kobayashi@ee.t.u-tokyo.ac.jp

**Abstract**

Two-dimensional (2D) systems, such as high-temperature superconductors, surface states of topological insulators, and layered materials, have been intensively studied using vacuum-ultraviolet (VUV) angle-resolved photoemission spectroscopy (ARPES). In semiconductor films (heterostructures), quantum well (QW) states arise due to electron/hole accumulations at the surface (interface). The quantized states due to quantum confinement can be observed by VUV-ARPES, while the periodic intensity modulations along the surface normal ($k_z$) direction of these quantized states are also observable by varying incident photon energy, resembling three-dimensional (3D) band dispersion. We have conducted soft X-ray (SX) ARPES measurements on thick and ultrathin III-V semiconductor InSb(001) films to investigate the electronic dimensionality reduction in semiconductor QWs. In addition to the dissipation of the $k_z$ dispersion, the SX-ARPES observations demonstrate the changes of the symmetry and periodicity of the Brillouin zone in the ultrathin film as 2D QW compared with these of the 3D bulk one, indicating the electronic dimensionality reduction of the 3D bulk band dispersion caused by the quantum confinement. The results provide a critical diagnosis using SX-ARPES for the dimensionality reduction in semiconductor QW structures.

## 1. Introduction

Quantum size effects have been widely used in practical semiconductor devices such as solar cells[1] and high-electron-mobility transistors[2,3]. In these devices, the quantum size effects derive from quantum confinement in quantum well (QW) structures and two-dimensional (2D) electron gas states at the interfaces. As to 2D systems, inherent low-dimensional systems different from conventional semiconductors have been intensively studied so far, e.g., high-temperature superconductors[4,5], non-trivial surface states of topological insulators[6,7], interface states in heterostructures[8], and layered materials[9,10]. In contrast to these intrinsic low-dimensional systems, the quantum confinement of valence or conduction electrons in semiconductor QW structures possibly reduces the dimensionality from three-dimensional (3D) to 2D.

The band structures of the low-dimensional systems have been studied by angle-resolved photoemission spectroscopy (ARPES) to unveil the electronic structures[11,12,13,14,15,16]. Commonly, ARPES with vacuum-ultraviolet (VUV) rays has been used to investigate the electronic band structures of these 2D materials because of the surface sensitivity (the probing depth of several Å), and high energy and momentum resolutions. The QW structure is one kind of quasi-2D system, and metal-based QWs such as $SrVO_3$[17,18] and Bi[19,20] ultrathin films have been studied by VUV-ARPES. Concerning semiconductor-based QW structures, while there have been few reports studying QW structures with ultrathin films by ARPES[21,22], QW states based on triangular potentials due to charge accumulations near the surfaces or interfaces have been intensively studied by ARPES so far[13,23,24,25]. In these triangular QW structures, the photon energy ($hv$) dependence of the quasi-2D QW states, which corresponds to the change of the surface normal momentum $k_z$ in a 3D crystal, shows the modulations of the ARPES intensities along the $k_z$ direction[13,23,26], like 3D $k_z$ dispersion. This is different from the dissipation of $k_z$ dispersion in 2D electronic structure[27]. Thus, it seems difficult to judge the dimensional reduction in the QW structures from $hv$ dependence on ARPES images.

In general, the 2D electronic band dispersion has a characteristic feature of dissipation of the $k_z$ dispersion compared with the 3D bulk dispersion. In addition to the non-dispersive $k_z$ band structure, the quantized band dispersion and the alternation of the symmetry from the 3D band dispersion due to the dimensional reduction are also characteristic 2D features for semiconductor QW structures. While the inherent 2D materials with layered structures are crystallized 2D, the quantum confinement in semiconductor QW makes the wave function 2D although the crystal structure is 3D (even with the ultrathin thicknesses). As shown in Fig. 1, the electronic dimensionality reduction due to the quantum confinement for the (001) orientation of a III-V semiconductor compound with the zinc-blende crystal structure will change the symmetry and the periodicity of the band dispersion as the same as these of the surface projected Brillouin zone (BZ).

To verify the dimensionality reduction in semiconductor QWs, we have performed soft X-ray (SX) ARPES on III-V semiconductor InSb thin films with different thicknesses, where one of the films is designed to form a QW structure due to the quantum confinement. SX-ARPES has been used to measure 3D crystals and interface states in heterostructures[28,29] because of its bulk sensitivity (the probing depth of several nm) and high $k_z$ resolution. In contrast to the $k_z$ broadening effects in VUV-ARPES[30], SX-ARPES has the advantage of checking the two dimensionality of the electronic structures by measuring the $k_z$ dispersion. In this paper, we have demonstrated the dimensionality reduction of the valence-band structure in the InSb QW ultrathin film using SX-ARPES.

## 2. Experimental

InSb films with 10 nm and 300 nm thickness were grown on p-GaAs(001) substrates by molecular beam epitaxy (MBE). To avoid surface contamination, the surfaces of the films were covered with amorphous Sb capping layers. Figure 2(a) shows the schematic sample structure; InSb/AlSb/AlAs/p-GaAs, from top to bottom. Considering previous studies[31,32] suggesting that InSb film with a thickness of 30 nm shows 2D features, the ultrathin InSb is designed to be a QW structure. Figure 2(b) shows the band alignment of the ultrathin film. Here, the AlSb layer acts as the barrier layer. Since the band gap of InSb is much narrower than that of AlSb and is located in that of AlSb, the quantum confinement in this film will quantize both the valence band (VB) and the conduction band (CB). During the MBE growth, the excellent crystallinity of the samples was confirmed by reflection high-energy electron diffraction (RHEED), as shown in Fig. 2(c). The SX-ARPES experiments were performed at the SX-ARPES end station[33] of the ADRESS beamline[34] at the Swiss Light Source. The Fermi-level positions of the ARPES spectra have been determined by measuring the Fermi cutoff of an Au foil in electrical contact with the samples. To expose the clean surface of the samples[35], the samples were annealed at around 300 ºC in the preparation chamber just before the SX-ARPES measurements to remove the amorphous Sb capping layer. The measurements were conducted under an ultrahigh vacuum below $10^{-9}$ Pa at a temperature of 15 K, varying the photon energy ($h\nu$) from 400 eV to 1100 eV. The total energy resolutions including the thermal broadening were between 50 meV and 200 meV depending on $h\nu$. The incident beam with circular polarization (*c*-polarization)[34] was used for the measurements.

### 3. Results and discussion

Figures 3(a) and 3(b) show the out-of-plane $k_{[001]}$-$k_{[\bar{1}10]}$ constant energy mappings (CEMs) of the thick and ultrathin InSb films at binding energy $E_B$ = 0.5 eV measured by a *c*-polarized ray. Here, $k_{[001]}$ is consistent with the surface normal momentum $k_z$. As shown in Fig. 3(a), the CEM pattern of the thick InSb film depends on $k_z$, and no non-dispersive band along the $k_z$ direction is observed in the CEM. These results indicate the observation of the bulk 3D band and provide evidence for the absence of surface 2D states in the present SX-ARPES data. On the other hand, as shown in Fig. 3(b), the CEM of the ultrathin InSb film, which is the second-derivative image of the raw CEM along the $k_{[\bar{1}10]}$ direction and symmetrized along the $k_{[\bar{1}10]}$ = 0 line, shows vertical line patterns along the $k_z$ direction and photoelectron intensity at the other points than the Γ points of 3D BZ which is different from the triangular-potential-based 2D systems[13], whose reason is mentioned later. By taking the second derivative of the cut of the CEM along the $k_{[-110]}$ direction, the peak of the cut becomes obvious. The observation indicates that the band dispersion along the $k_{[001]}$ direction disappears, consistent with the 2D electronic states due to the quantum confinement in the QW structure.

Figures 3(c)-3(e) and 3(f)-3(h) are ARPES images of the thick and ultrathin InSb films taken at $h\nu$ = 790, 850, and 910 eV, respectively. The cuts at these $h\nu$s are represented by yellow curves in Fig. 3(a). In the out-of-plane CEM of the thick InSb film in Fig. 3(a), the cuts taken at $h\nu$ = 790 and 910 eV contain the Γ point. The out-of-plane CEM in Fig. 3(a) and the ARPES images in Figs. 3(c)-3(e) reflect the symmetry of the 3D bulk BZ for the zinc blende structure, and these bands are identified as the light-hole (LH), heavy-hole (HH), and split-off (SO) bands of InSb[36]. This VB structure is a typical band structure of III-V semiconductors[35,37]. These observations demonstrate that the SX-ARPES spectra reflect the 3D band dispersion of the thick InSb film. The valence band maximum (VBM) is estimated to be located at $E_B$ = 0.09 ± 0.03 eV. Although InSb is n-type, the reason for the position of $E_F$ in the band gap would be the pinning effect[38].

Figures 3(f)-3(h) show the series of the ARPES images of the ultrathin InSb film, where the red markers represent the peak positions of the momentum distribution curves. Unlike the thick InSb film, the band dispersion of the ultrathin InSb film is independent of $hv$. Since the surface states are not observed in the present SX-ARPES measurements on the thick InSb film, it is probable that any surface states hardly contribute to the observed band structure of the ultrathin InSb film. Since no Al core-level spectra have been observed with $hv$ up to 1100 eV in both the thick and ultrathin InSb films (not shown), the ARPES spectra of the ultrathin InSb film do not originate from the AlSb buffer layer. The top of the observed band of the ultrathin InSb film is located at $E_B = 0.24 \pm 0.05$ eV. To identify the observed band, the simple Schrödinger equation of QW is used:

$$\left(-\frac{\hbar^2}{2m^*}\frac{d^2}{dx^2} + V\right)\psi(x) = E\psi(x)$$

where $m^*$ and $V$ represent the effective mass and potential barrier from the edge of VB (conduction band (CB)) of InSb, respectively. $V$ is set to ∞ in vacuum, 0.98 eV for electrons in AlSb, and 0.45 eV for holes in AlSb, and $m^*$ of electrons, LHs, and HHs are set to $0.015m_0$, $0.021\ m_0$, and $0.39\ m_0$[39], respectively. Here, $m_0$ represents the free-electron mass. Figure 2(b) shows the quantized states obtained by solving the Schrödinger equation. Based on the ARPES spectra, the top of the observed band of the ultrathin InSb film is $0.15 \pm 0.08$ eV below the VBM of the thick InSb film. Since the first quantized state of the LH band is 122 meV below the VBM of the 3D InSb in Fig. 2(b), the top of the observed band of the ultrathin InSb film would be the first quantized state of the LH band.

In the out-of-plane CEM of the thick InSb film (not shown here), the cut taken at $hv$ = 695 eV also contains the Γ point at $k_{[\bar{1}10]} = 0$. Then, the Γ-K-X symmetry line is precisely determined from the in-plane $k_{[\bar{1}10]}$-$k_{[\bar{1}\bar{1}0]}$ CEM at $E_B = 0.5$ eV taken at $hv$ = 695 eV with $c$-polarization, as shown in Fig. 4(a). As well as the out-of-plane CEM in Fig. 3(a), the in-plane CEM reflects the symmetry of the 3D bulk BZ for the zinc blende structure. Figure 4(b) shows the ARPES image along the Γ-K-X symmetry line (green dashed line in Fig. 4(a)) taken with a $c$-polarized ray and the HH, LH, and SO bands as well as Fig. 3(c).

Since the $k_z$ dispersion in the ultrathin film is disappeared as shown in Figs. 3(f)-3(h), we have chosen $hv$ = 642 eV for the following ARPES measurements due to the large intensity of photoelectrons. Figure 4(c) shows the in-plane $k_{[\bar{1}10]}$-$k_{[\bar{1}\bar{1}0]}$ CEM of the ultrathin InSb film at $E_B$ = 0.5 eV is taken at $hv$ = 642 eV with $c$-polarization. While the in-plane CEM of the bulk InSb film shows finite intensities around the Γ points of the 3D BZ, the in-plane CEM of the InSb QW shows finite intensities around the X points of the 3D BZ corresponding to the Γ' points of the 2D BZ. Thus, the in-plane CEM pattern reflects the same symmetry of the 2D BZ, different from the 3D BZ. Figure 4(d) shows the ARPES image along the Γ'-X'-Γ' symmetry line (orange dashed line in Fig. 4(c), see Fig. 1) taken with a $c$-polarized ray. Since the top of the observed band is located near $E_F$, the band would be the quantized HH bands. As well as the in-plane CEM, the band dispersion of the ultrathin InSb film is different from that of the thick InSb film. Compared with the bulk band dispersion (Fig. 4(b)), the band dispersion of the ultrathin InSb shows the maximum values at $k_{[\bar{1}\bar{1}0]} = 0$ and 1 ($2\sqrt{2}\pi/a$), which correspond with the Γ and X points in the 3D BZ, respectively. Note that $k_{[\bar{1}\bar{1}0]}$ is equivalent to $k_{[\bar{1}10]}$ due to the crystal symmetry. This result indicates that the periodicity of the band dispersion becomes half in the ultrathin InSb film. It should be noted that the crystal structure of the ultrathin InSb film is the same as that of the thick InSb film even though the thickness in ultrathin InSb film is thin enough to induce the quantum confinement. Because only the dimensionality of the electron wave function is reduced by the quantum confinement while maintaining its crystal structure, this result provides the experimental evidence for the electronic dimensionality reduction in semiconductor QW structures. Since the in-plane 2D electronic states have not been observed in the triangle-potential QWs[13,23,24,25], the electronic states of the ultrathin InSb film is purely 2D different from the triangle-potential QWs, leading to the non-dispersive out-of-plane electronic states in Fig. 3(b).

## 4. Conclusion

To verify the dimensionality reduction in semiconductor QW, we have conducted SX-ARPES measurements on thick and ultrathin III-V semiconductor InSb films, the latter of which is thin enough to include the quantum confinement. The observations in thick InSb film demonstrate that SX-ARPES enables us to observe the entire 3D bulk band dispersion of the thick InSb film. In contrast, the SX-ARPES results on the InSb QW film show the dissipation of the $k_z$ dispersion and the changes in the symmetry and periodicity of BZ compared with those of the thick InSb one. The results indicate that the dimensionality reduction of the 3D band dispersion occurs due to the quantum confinement in the QW structure. The results provide a critical diagnosis using SX-ARPES for the electronic dimensionality reduction in semiconductor QW structures. This suggests the importance of SX-APRES in the evaluation of semiconductor QWs.

**Acknowledgment**

The authors thank A. Fujimori for enlightening discussion. This work was supported by Grants-in-Aid for Scientific Research (19K1961, 20H05650, 23K17324.) and CREST (JPMJCR1777) of Japan Science and Technology. This work was also supported by the Spintronics Research Network of Japan (Spin-RNJ).

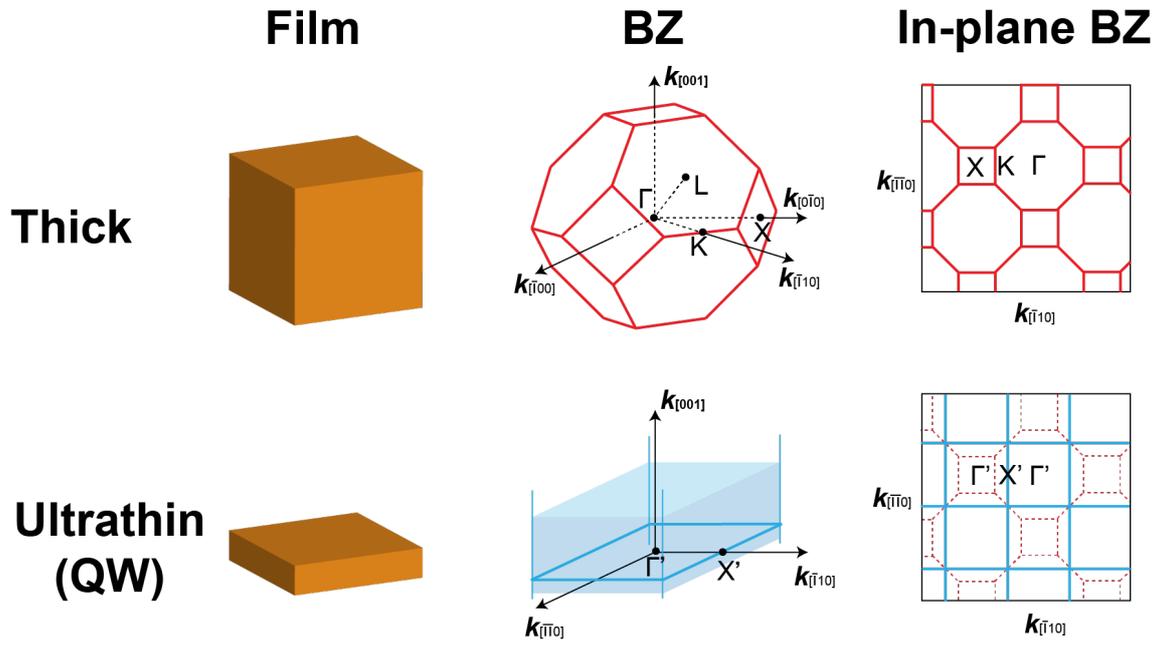

FIG. 1. Concept of the electronic dimensionality reduction from a thick semiconductor film to an ultrathin film. The upper and lower rows represent the cases of thick and ultrathin films, respectively. The schematic pictures of films, BZs, and in-plane BZs are described at the left, middle, and right columns, respectively.

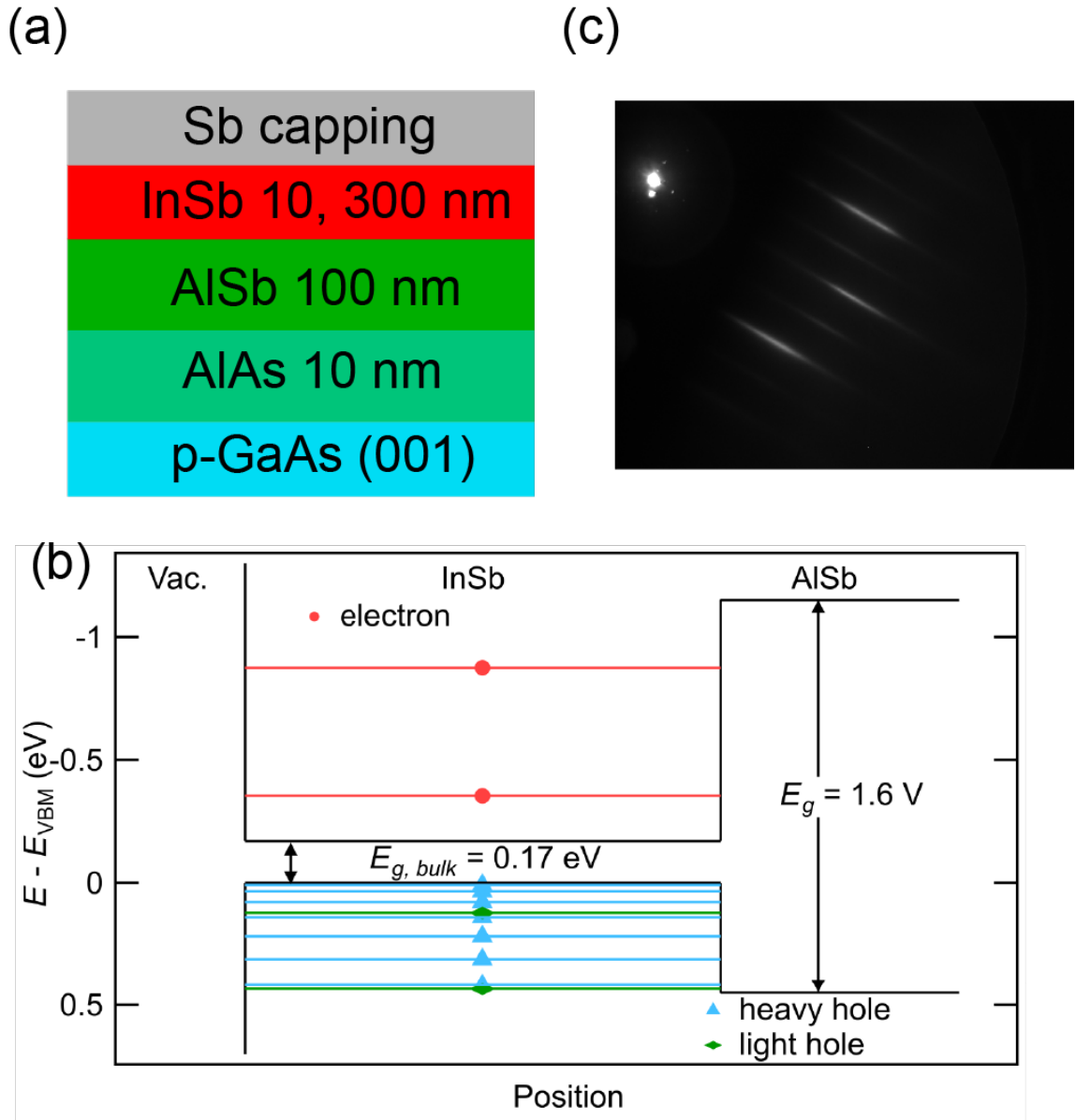

FIG. 2. Sample structure of InSb films. (a) Structure of the InSb films with the thickness of 10 and 300 nm. (b) Flat band lineup of vacuum (Vac.)/InSb/AlSb, where $E_{g,Bulk}$ and $E_g$ represent the band gap of bulk InSb and AlSb, respectively. $E_{VBM}$ denotes the top of the valence band of InSb. (c) RHEED pattern along the [110] direction of bulk InSb.

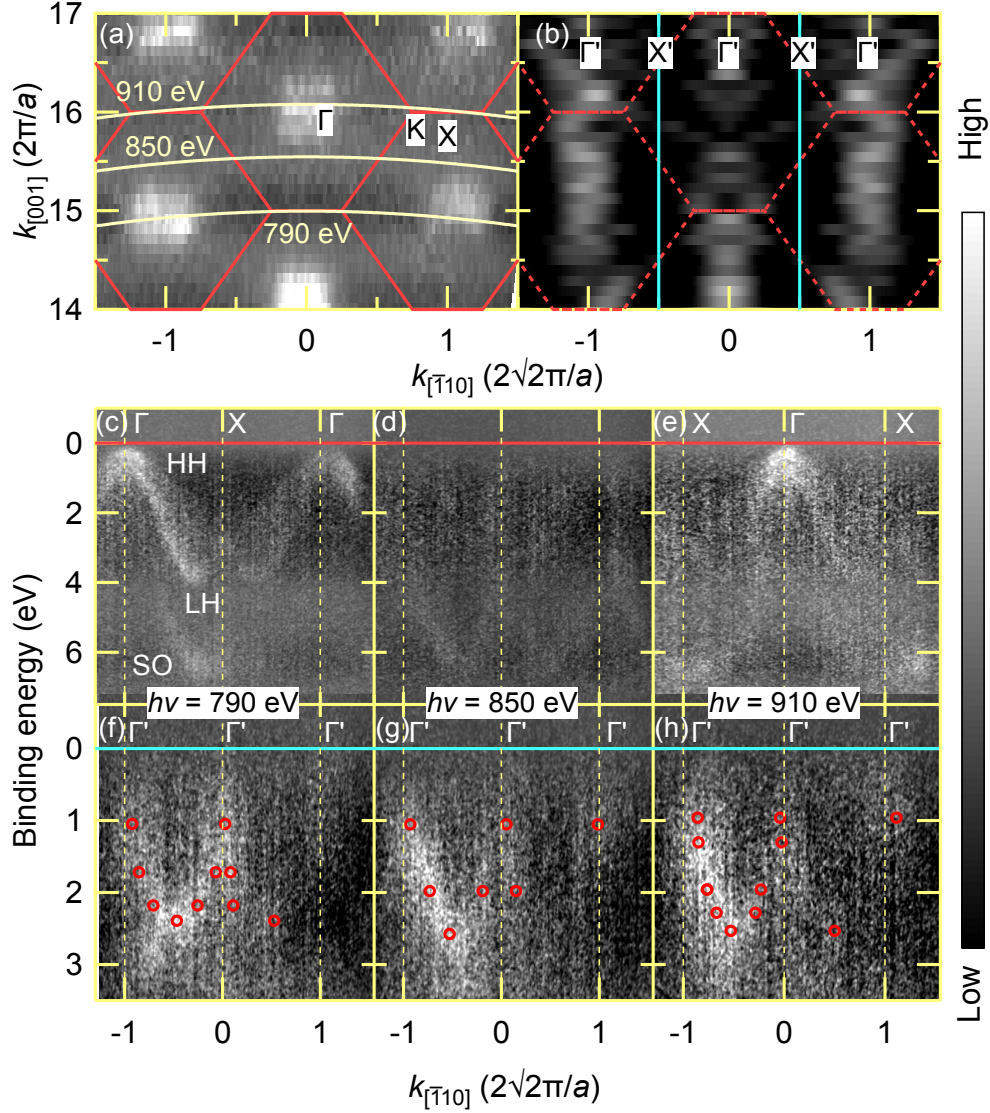

FIG. 3. Out-of-plane CEMs of thick (a) and ultrathin (b) InSb films at $E_B$ = 0.5 eV in the $k_{[001]}$-$k_{[\bar{1}10]}$ plane taken with *c*-polarization, where the red solid and blue lines represent the 3D and 2D BZ boundaries, respectively. As for (b), the CEM is the second-derivative image of the raw CEM along the $k_{[\bar{1}10]}$ direction and symmetrized along the $k_{[\bar{1}10]}$ = 0 line to emphasize the peak position. (c, d, e) ARPES images of the thick InSb film along the $k_{[\bar{1}10]}$ direction taken at $hv$ = 790, 850, and 910 eV, respectively. (f, g, h) ARPES images of the ultrathin InSb film along the $k_{[\bar{1}10]}$ direction taken at $hv$ = 790, 850, and 910 eV, respectively. The cuts taken at the photon energies are corresponding to the curves in Fig. 3 (a). Red markers represent the peak position of the momentum distribution curves.

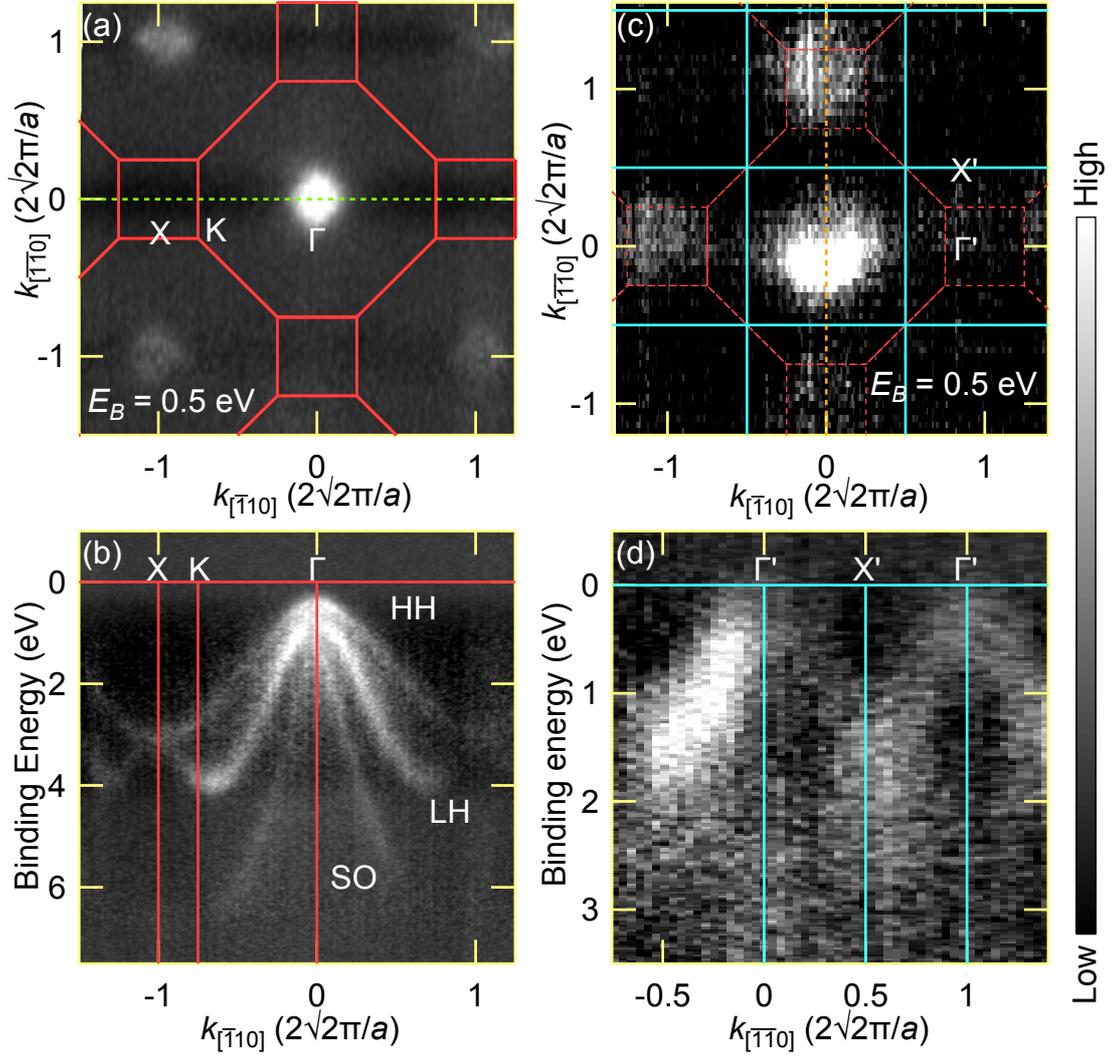

FIG. 4. Change of the periodicity due to the dimensionality reduction. (a) In-plane CEM of the thick InSb film at $E_B = 0.5$ eV in the $k_{[\bar{1}10]}$-$k_{[\bar{1}\bar{1}0]}$ plane taken at $h\nu = 695$ eV. The red solid and green dashed lines represent the 3D BZ and the Γ-K-X line, respectively. (b) ARPES image along the Γ-K-X line (green dashed line in Fig. 4(a)) taken at $h\nu = 695$ eV. (c) In-plane CEM of the ultrathin InSb film at $E_B = 0.5$ eV in the $k_{[\bar{1}10]}$-$k_{[\bar{1}\bar{1}0]}$ plane taken at $h\nu = 642$ eV. The blue, red, and orange dashed lines represent the 2D BZ, 3D BZ, and the Γ-K-Γ line, respectively. (d) ARPES image along the Γ'-K'-Γ' line (orange dashed line in Fig. 4(c)) taken at $h\nu = 642$ eV. Here, the CEM and ARPES images have been taken with $c$-polarization.

14